\documentclass[11pt]{article}
\usepackage[pctex32]{graphics}

\def\ben{\begin{equation}}
\def\een{\end{equation}}

\def\nn{\nonumber} \def\bd{\begin{document}} \def\ed{\end{document}}
\def\ds{\documentstyle} \let\fr=\frac \let\bl=\bigl \let\br=\bigr
\let\Br=\Bigr \let\Bl=\Bigl
\let\bm=\bibitem
\let\na=\nabla
\let\pa=\partial \let\ov=\overline
\newcommand{\be}{\begin{equation}}
\newcommand{\ee}{\end{equation}}
\def\ba{\begin{array}}
\def\ea{\end{array}}
\def\ft#1#2{{\textstyle{\frac{\scriptstyle #1}{\scriptstyle #2} } }}
\def\fft#1#2{{\frac{#1}{#2}}}
\def\del{\partial}
\def\vp{\varphi}
\def\sst#1{{\scriptscriptstyle #1}}
\def\oneone{\rlap 1\mkern4mu{\rm l}}
\def\td{\tilde}
\def\wtd{\widetilde}
\def\ie{{\it i.e.\ }}
\def\dalemb#1#2{{\vbox{\hrule height .#2pt
        \hbox{\vrule width.#2pt height#1pt \kern#1pt
                \vrule width.#2pt}
        \hrule height.#2pt}}}
\def\square{\mathord{\dalemb{6.8}{7}\hbox{\hskip1pt}}}
\newcommand{\ho}[1]{$\, ^{#1}$}
\newcommand{\hoch}[1]{$\, ^{#1}$}
\newcommand{\bea}{\setlength\arraycolsep{2pt} \begin{eqnarray}}
\newcommand{\eea}{\end{eqnarray}}
\newcommand{\ra}{\rightarrow}
\newcommand{\lra}{\longrightarrow}
\newcommand{\Lra}{\Leftrightarrow}
\newcommand{\bp}{\tilde \beta^\prime}
\newcommand{\tr}{{\rm tr} }
\newcommand{\Tr}{{\rm Tr} }
\def\0{{\sst{(0)}}}
\def\1{{\sst{(1)}}}
\def\2{{\sst{(2)}}}
\def\3{{\sst{(3)}}}
\def\4{{\sst{(4)}}}
\def\5{{\sst{(5)}}}
\def\6{{\sst{(6)}}}
\def\7{{\sst{(7)}}}
\def\8{{\sst{(8)}}}
\def\m{{\sst{(m)}}}
\def\n{{\sst{(n)}}}
\def\cA{{{\cal A}}}
\def\cB{{{\cal B}}}
\def\cF{{{\cal F}}}
\def\cG{{{\cal G}}}
\def\cH{{{\cal H}}}
\def\tV{\widetilde V}
\def\tW{\widetilde W}
\def\tH{\widetilde H}
\def\tE{\widetilde E}
\def\tF{\widetilde F}
\def\tA{\widetilde A}
\def\im{{{\rm i}}}
\def\tY{{{\wtd Y}}}
\def\ep{{\epsilon}}
\def\vep{{\varepsilon}}
\def\bD{{{\bar D}}}
\def\R{{{\mathbb R}}}
\def\C{{{\mathbb C}}}
\def\H{{{\mathbb H}}}
\def\CP{{{\mathbb C}{\mathbb P}}}
\def\RP{{{\mathbb R}{\mathbb P}}}
\def\Z{{{\mathbb Z}}}
\def\bA{{{\mathbb A}}}
\def\bB{{{\mathbb B}}}
\def\bC{{{\mathbb C}}}
\def\bD{{{\mathbb D}}}
\def\bE{{{\mathbb E}}}
\def\bZ{{{\mathbb Z}}}
\def\Re{{{\frak{Re}}}}
\def\Im{{{\frak{Im}}}}
\def\cosec{{\,\hbox{cosec}\,}}
\def\Gm{{\Gamma_{\!\! -}}}
\def\Gp{{\Gamma_{\!\! +}}}
\def\stan{{standard }}
\def\nonstan{{supernumerary }}
\def\p{{\partial}}
\def\kdel#1{{\fft{\del}{\del#1}}}

\def\bog{{Bogomolny }}
\def\om{{\varpi}}

\newcommand{\nnr}{\nonumber \\}
\newcommand{\pd}{\partial}
\newcommand{\ud}{\textrm{d}}
\newcommand{\dTH}{T^{\prime \, 0}_\textrm{H}}
\newcommand{\dOi}{\Omega^{\prime \, 0}_i}
\newcommand{\bx}{{\bf x}}

\begin{document}

\vspace{5mm}
\begin{center}
{\Large \bf Generalized uncertainty principle, quantum gravity and
Ho\v{r}ava-Lifshitz gravity } \vspace{12mm}

{\large   Yun Soo Myung \footnote{e-mail
 address: ysmyung@inje.ac.kr}}
 \\
\vspace{10mm} {\em Institute of Basic Science and School of
Computer Aided Science \\ Inje University, Gimhae 621-749, Korea}
\end{center}

\begin{center}

\underline{Abstract}
\end{center}

  We investigate a close connection between generalized uncertainty
  principle (GUP) and deformed Ho\v{r}ava-Lifshitz (HL) gravity.
The GUP commutation relations correspond to the UV-quantum theory,
while the canonical commutation relations represent the IR-quantum
theory. Inspired by this UV/IR quantum mechanics,  we obtain the
GUP-corrected graviton propagator by introducing UV-momentum
$p_i=p_{0i}(1+\beta p_{0}^2)$ and compare this with tensor
propagators in the HL gravity. Two are the same up to
$p_0^4$-order. \vspace{15pt}

\thispagestyle{empty}





\newpage
\section{Introduction}
Recently Ho\v{r}ava has proposed a renormalizable theory of
gravity at a Lifshitz point~\cite{ho1},  which  may be regarded as
a UV complete candidate for general relativity. At short distances
the theory of $z=3$ Ho\v{r}ava-Lifshitz (HL) gravity describes
interacting nonrelativistic gravitons and is supposed to be power
counting renormalizable in (1+3) dimensions. Recently, the HL
gravity theory has been intensively investigated
in~\cite{ho2,Vi,ho3,VW,Kl1,Ni,Na,Iz,Vo,SVW1,CH,CHZ,Nis,OR,KLM1,Ko,LP,CNPS,SVW2,Ca,Koch,Sa,My,GKS,BPS,Kob,BS}.
The equations of motion were derived for $z=3$ HL
gravity~\cite{KK,LMP}, and its black hole solution was first found
in asymptotically anti-de Sitter spacetimes~\cite{LMP} and black
hole in asymptotically flat spacetimes~\cite{KS}.

It seems that the GUP-corrected Schwarzschild black hole is
closely related to black holes in the deformed Ho\v{r}ava-Lifshitz
gravity~\cite{Myung,Myungent}. Also,  the GUP provides naturally a
UV cutoff to the local quantum field theory as quantum gravity
effects~\cite{CMOK,KLM2}.

On the other hand,  one of main ingredients for studying quantum
gravity is the GUP, which has been argued from various approaches
to quantum gravity and black hole physics~\cite{gup}. Certain
effects of quantum gravity are universal and thus, influence
almost any system with a well-defined Hamiltonian~\cite{DV}.
 The GUP satisfies the
modified Heisenberg algebra~\cite{KMM} \be \label{UVCR}
[x_i,p_j]=i\hbar\Big(\delta_{ij}+\beta p^2 \delta_{ij}+2\beta
p_ip_j\Big),~[x_i,x_j]=[p_i,p_j]=0\ee where $p_i$ is considered as
the momentum at high energies and thus, it can be interpreted to
be the UV-commutation relations. Here $p^2=p_ip_i$. In this case,
the minimal length
 which follows from these relations is given by
 \be
 \delta x_{\rm min}=\hbar \sqrt{5\beta}.\ee
On the other hand, introducing IR-canonical variable  $p_{0i}$
with $x_i=x_{0i}$ through the replacement \be
\label{RPCR}p_i=p_{0i}\Big(1+\beta p_0^2\Big),\ee these variables
satisfy canonical commutation relations \be \label{IRCR}
[x_{0i},p_{0j}]=i\hbar\delta_{ij},~[x_{0i},x_{0j}]=[p_{0i},p_{0j}]=0.
\ee Here  $p_{0i}$ is considered as the momentum at low energies
with $p_0^2=p_{0i}p_{0i}$. It is easy to show that  Eq.
(\ref{UVCR}) is satisfied to linear-order $\beta$ when using Eq.
(\ref{IRCR}). Hence, the replacement (\ref{RPCR}) could be used as
an important low-energy window to investigate quantum gravity
phenomenology up to linear-order $\beta$.

 It was known for deformed HL gravity that the
UV-propagator  for tensor modes $t_{ij}$ take a complicated form
Eq. (\ref{tenprp}), including up to $p_{0}^6$-term from the Cotton
bilinear term $C_{ij}C_{ij}$. We have explored a connection
between the GUP commutator  and the deformed HL
gravity~\cite{Myunggup}. Explicitly, we have replaced a
relativistic cutoff function $\mathcal{K}(\frac{p^2}{\Lambda^2})$
by a non-relativistic density function ${\cal D}_{D}(\beta
\vec{p}^2)$ to derive GUP-corrected graviton propagators. These
were compared to (\ref{tenprp}). It was pointed out that two are
{\it qualitatively similar}, but the $p^5$-term arisen from the
crossed term of Cotton and Ricci tensors did not appear in the
GUP-corrected propagators. Also, it was  unclear why the $D=2$
GUP-corrected tensor propagator (not
 the $D=3$ GUP-corrected propagator)
 is  similar to  the UV-propagator  derived from the $z=3$ HL gravity.

 In this work, we investigate a close connection between
GUP and deformed HL gravity.  At high energies, we assume that the
UV-propagator takes the conventional form $G_{\rm UV}(\varpi,p^2)$
in Eq. (\ref{UVprop}), whereas at low energies,  the IR-propagator
takes the conventional form $G_{\rm IR}(\varpi,p_0^2)$ in Eq.
(\ref{IRprop}). It is very important to understand how the
UV-propagator is related to the IR-propagator in the
non-relativistic gravity theory.
 We find a
GUP-corrected graviton propagator by applying (\ref{RPCR}) to
$G_{\rm UV}(\varpi,p^2)$ and compare it with the UV-tensor
propagator
 (\ref{tenprp}) in the HL gravity.  Two are {\it the same} up to $p_0^4$-order, although
the $p_0^5$-term arisen from a crossed term of  Cotton tensor and
Ricci tensor is still  missed in the GUP-corrected graviton
propagator. This indicates that  a power-counting renormalizable
theory of the HL gravity is closely related to the GUP.

 \section{$z=3$ HL gravity}
Introducing the ADM formalism where the metric is parameterized
\be ds_{ADM}^2= - N^2  dt^2 + g_{ij} \Big(dx^i - N^i dt\Big)
\Big(dx^j - N^j dt\Big)\,, \ee
the Einstein-Hilbert action can be expressed as
\be \label{Eins} S_{EH} = \fft{1}{16\pi G} \int d^4x \sqrt{g} N
\Big[K_{ij} K^{ij} - K^2 + R - 2\Lambda\Big], \ee
where $G$ is Newton's constant and extrinsic curvature $K_{ij}$
takes the form
\be K_{ij} = \fft{1}{2N} \Big(\dot g_{ij} - \nabla_i N_j -
\nabla_j N_i\Big)\,. \ee
Here, a dot denotes a derivative with respect to $t$. An action of
the non-relativistic renormalizable gravitational theory  is given
by~\cite{ho1} \be S_{HL}=\int dtd^3x \Big[{\cal L}_K + {\cal
L}_V\Big],  \label{action1} \ee where the kinetic terms are given
by \be {\cal L}_K =\frac{2}{\kappa^2}\sqrt{g} N K_{ij}{\cal
G}^{ijkl}K_{kl}= \frac{2}{\kappa^2}\sqrt{g}
N\Big(K_{ij}K^{ij}-\lambda K^2\Big), \ee with the DeWitt metric
 \be {\cal G}^{ijkl}=
\frac{1}{2}\Big(g^{ik}g^{jl}-g^{il}g^{jk}\Big)-\lambda
g^{ij}g^{kl} \ee
 and its inverse metric
 \be {\cal
G}_{ijkl}=\frac{1}{2}\Big(g_{ik}g_{jl}-g_{il}g_{jk}\Big)-\frac{\lambda}{3\lambda-1}g_{ij}g_{kl}.\ee

The potential terms is determined by the detailed balance
condition  as \bea {\cal L}_V=-\frac{\kappa^2}{2}\sqrt{g}N
E^{ij}{\cal G}_{ijkl}E^{kl}&=&
\sqrt{g}N\Bigg\{\frac{\kappa^2\mu^2}{8(1-3\lambda)}\Big(\frac{1-4\lambda}{4}R^2
+\Lambda_WR-3\Lambda_W^2\Big)\nn \\
 &-&\frac{\kappa^2}{2\eta^4} \left(C_{ij}
-\frac{\mu \eta^2}{2}R_{ij}\right) \left(C^{ij} -\frac{\mu
\eta^2}{2}R^{ij}\right) \Bigg\}.\label{action2} \eea Here the $E$
tensor is defined by \be E^{ij}=\frac{1}{\eta^2}
C^{ij}-\frac{\mu}{2} \Big(R^{ij}-\frac{R}{2}
g^{ij}+\Lambda_Wg^{ij}\Big) \ee with the Cotton tensor $C_{ij}$
\be
C^{ij}=\frac{\epsilon^{ik\ell}}{\sqrt{g}}\nabla_k\left(R^j{}_\ell
-\frac14R\delta_\ell^j\right).\label{def.K.C} \ee  Explicitly,
$E_{ij}$ could be derived  from the Euclidean topologically
massive gravity \be E^{ij}=\frac{1}{\sqrt{g}} \frac{\delta
W_{TMG}}{\delta g_{ij}} \ee with \be W_{TMG}=\frac{1}{\eta^2} \int
d^3 x \epsilon^{ikl}\Big(\Gamma^m_{il}\partial_j
\Gamma^l_{km}+\frac{2}{3} \Gamma^n_{il} \Gamma^l_{jm}
\Gamma^m_{kn} \Big)- \mu \int d^3x \sqrt{g}(R-2\Lambda_W), \ee
where $\epsilon^{ikl}$ is a tensor density with
$\epsilon^{123}=1$.

In the IR limit,  comparing ${\cal L}_0$ with Eq.(\ref{Eins}) of
general relativity, the speed of light, Newton's constant and the
cosmological constant are given by
\be c=\fft{\kappa^2\mu}{4}
\sqrt{\fft{\Lambda_W}{1-3\lambda}}\,,\qquad
G=\fft{\kappa^2}{32\pi\,c}\,,\qquad \Lambda_{\rm cc}=\ft32
\Lambda_W\,.\label{cg} \ee The equations of motion were derived in
\cite{KK} and \cite{LMP}. We would like to mention that the IR
vacuum of this theory is anti-de Sitter (AdS$_4$) spacetimes.
Hence, it is interesting to take a limit of the theory, which may
lead to  a Minkowski vacuum in the IR sector. To this end, one may
deform the theory by introducing ``$\mu^4R$" $(\tilde{{\cal
L}}_V={\cal L}_V+\sqrt{g}N \mu^4R)$ and then, take the $\Lambda_W
\to 0$ limit~\cite{KS}. We call this the deformed HL gravity
without detailed balance condition. This does not alter the UV
properties of the theory, while it changes the IR properties. That
is, there exists a Minkowski vacuum, instead of an AdS vacuum. In
the IR limit, the speed of light and Newton's constant are given
by
\be c^2=\fft{\kappa^2\mu^4}{2},~ G=\fft{\kappa^2}{32\pi\,c},
~\lambda=1.\label{kh} \ee The deformed HL gravity has an important
parameter~\cite{KS} \be
\omega=\frac{8\mu^2(3\lambda-1)}{\kappa^2},\ee which takes the
form for $\lambda=1$ \be \omega=\frac{16\mu^2}{\kappa^2}.\ee
Actually, $\frac{1}{2\omega}$ plays the role of a charge in  the
Kehagias-Sfetsos (KS) black hole with $\lambda=1$ and
$K_{ij}=C_{ij}=0$~\cite{Myung} derived from the Lagrangian \be
\label{kslag} \tilde{{\cal L}}^{\lambda=1}_V=\sqrt{g}N \mu^4\Big(R
+\frac{3}{4\omega}R^2-\frac{2}{\omega}R_{ij}R_{ij}\Big) .\ee and a
spherically symmetric metric ansatz. Furthermore, it was shown
that  the entropy of KS black hole could be  explained from the
entropy of GUP-corrected Schwarzschild black hole when making a
connection of $\beta \to \frac{1}{\omega}$~\cite{Myungent}.
\section{GUP-quantum mechanics}
A meaningful prediction of various theories of quantum gravity
(string theory) and black holes is the presence of a minimum
measurable length or a maximum observable momentum. This has
provided the generalized uncertainty principle which modifies
commutation relations shown by Eq. (\ref{UVCR}). A universal
quantum gravity correction to the Hamiltonian is given by
\begin{eqnarray} \label{uvhamil1} {\cal
H}_{UV}&=&\frac{p^2}{2m}+V(x_i)=\frac{p_{0}^2}{2m}+
V(x_{0i})+\frac{\beta}{m}p_0^4+\frac{\beta^2}{2m}p_{0}^6  \\
\label{uvhamil2} &\equiv& {\cal H}_{IR}+{\cal H}_1
\end{eqnarray}
 with
\be {\cal H}_{IR}=\frac{p_{0}^2}{2m}+ V(x_{0i}),~~{\cal
H}_1=\frac{\beta}{m}p_0^4+\frac{\beta^2}{2m}p_{0}^6 .\ee
 We note that  Eq. (\ref{uvhamil2}) may  be used for a
perturbation study with $p_0=-i\hbar d/dx_{0i}$. We see that any
system with a well-defined quantum (or even classical) Hamiltonian
${\cal H}_{IR}$, is perturbed by ${\cal H}_1$ near the Planck
scale. In this sense, the quantum gravity effects are in  some
sense universal. Some examples were performed
in~\cite{DV,Sil,DVp,ADV}. It turned out that the corrections could
be interpreted in two ways when considering linear-order
perturbation  ${\cal H}_1=\frac{\beta}{m}p_0^4$: either that for
$\beta=\beta_0l_{\rm Pl}^2/2\hbar^2$ with $\beta_0 \sim 1$, they
are exceedingly small, beyond the reach of current experiments or
that they predict upper bounds on the quantum gravity parameter
$\beta_0 \le 10^{34}$ for the Lamb shift.

\subsection{Tensor modes for deformed $z=3$ HL gravity} The field equation for tensor modes propagating
on the Minkowski spacetimes is given by~\cite{My} \be
\ddot{t}_{ij}-\frac{\mu^4\kappa^2}{2} \bigtriangleup t_{ij}
+\frac{\mu^2\kappa^4}{16}\bigtriangleup^2t_{ij}-\frac{\mu\kappa^4}{4\eta^2}
\epsilon_{ilm}\partial^l\bigtriangleup^2t_j~^m-
\frac{\kappa^4}{4\eta^4} \bigtriangleup^3 t_{ij}=T_{ij} \ee with
external source $T_{ij}$ and the Laplacian
$\bigtriangleup=\partial_i^2\to-p_0^2$. We could not obtain
 the covariant propagator because of the presence of $\epsilon$-term.
  Assuming a massless graviton propagation along the
$x^3$-direction with $p_{0i}=(0,0,p_3)$, then the $t_{ij}$ can be
expressed in terms of polarization components as~\cite{BS} \be
t_{ij}=\left(%
\begin{array}{ccc}
  t_+ & t_\times & 0 \\
  t_\times & -t_+ & 0 \\
  0 &0  & 0 \\
\end{array}%
\right). \ee Using this parametrization, we find two coupled
equations for different polarizations \bea \ddot{t}_+
-\frac{\mu^4\kappa^2}{2} \bigtriangleup t_+
+\frac{\kappa^4\mu^2}{16}\bigtriangleup^2t_+
+\frac{\kappa^4\mu}{4\eta^2}\partial_3\bigtriangleup^2t_\times-\frac{\kappa^4}{4\eta^4}\bigtriangleup^3t_+=T_{+},
\\
\ddot{t}_\times-\frac{\mu^4\kappa^2}{2} \bigtriangleup
t_\times+\frac{\kappa^4\mu^2}{16}\bigtriangleup^2t_\times
-\frac{\kappa^4\mu}{4\eta^2}\partial_3\bigtriangleup^2t_+-\frac{\kappa^4}{4\eta^4}\bigtriangleup^3t_\times=T_{\times}.
\eea In order to find two independent components, we introduce the
left-right base defined by \be
t_{L/R}=\frac{1}{\sqrt{2}}\Big(t_+\pm it_\times\Big) \ee where
$t_L(t_R)$ represent the left (right)-handed modes.  After
Fourier-transformation, we find two decoupled equations \bea
-\varpi^2{t}_L+ c^2p_0^2 t_L +\frac{\kappa^4\mu^2}{16}(p_0^2)^2t_L
-\frac{\kappa^4\mu}{4\eta^2}p_3(p_0^2)^2t_L+\frac{\kappa^4}{4\eta^4}(p_0^2)^3t_L=
T_L,
\\
-\varpi^2{t}_R +c^2p_0^2 t_R +\frac{\kappa^4\mu^2}{16}(p_0^2)^2t_R
+\frac{\kappa^4\mu}{4\eta^2}p_3(p_0^2)^2t_R+\frac{\kappa^4}{4\eta^4}(p_0^2)^3t_R=
T_R. \eea We have UV-tensor propagators with
$\omega=16\mu^2/\kappa^2$ \be
\label{tenprp}t_{L/R}=-\frac{T_{L/R}}{ \varpi^2-c^2\Big(p_0^2
+\frac{2}{\omega}p_0^4 \mp
\frac{8}{\eta^2\mu\omega}p_3p_0^4+\frac{128}{\eta^4\kappa^2\omega^2}p_0^6\Big)}.\ee
We note that the left-handed mode is not allowed because it may
give rise to ghost ($-\frac{8c^2}{\eta^2\mu\omega}p_3p_0^4$),
while the right-handed mode is allowed because there is no ghost
($\frac{8c^2}{\eta^2\mu\omega}p_3p_0^4$). At this stage, we
mention that $p_0(=\sqrt{p_{0i}p_{0i}})$ is a magnitude of
momentum $p_{0i}$
 but not a time component $\varpi$.

Finally, we find UV-propagators in the four dimensional frame with
$p^\mu=(\varpi,0,0,p_3)$ as \be t_{L/R}=-\frac{T_{L/R}}{
\varpi^2-c^2\Big(p_3^2+\frac{2}{\omega}p_3^4 \mp
\frac{8}{\eta^2\mu\omega}p_3^5+\frac{128}{\eta^4\kappa^2\omega^2}p_3^6\Big)}.\ee

\subsection{GUP-corrected propagator}
It is known for deformed HL gravity that the UV-propagator  for
tensor modes $t_{ij}$ take a complicated form  shown in  Eq.
(\ref{tenprp}),  including up to $p_{0}^6$-term from the Cotton
bilinear term $C_{ij}C_{ij}$.

 At high energies, we assume that
the UV-propagator takes the conventional form  \be \label{UVprop}
G_{\rm UV}(\varpi,p^2)=\frac{1}{\varpi^2-c^2p^2},\ee whereas at
low energies,  the IR-propagator takes the conventional form  \be
\label{IRprop}G_{\rm
IR}(\varpi,p^2_{0})=\frac{1}{\varpi^2-c^2p_{0}^2}.\ee Considering
(\ref{RPCR}), the UV-propagator (\ref{UVprop})  takes the form \be
G_{\rm UV}(\varpi,p_{0}^2)=\frac{1}{\varpi^2-c^2\Big(
p_{0}^2+2\beta p_{0}^4+\beta^2p_{0}^6\Big)}. \ee The GUP-corrected
tensor propagator is determined by  \be \label{gupcoprop}
t^{GUP}_{ij} = -G_{\rm
UV}(\varpi,p_0^2)T_{ij}=-\frac{T_{ij}}{\varpi^2-c^2\Big(
p_{0}^2+2\beta p_{0}^4+\beta^2p_{0}^6\Big)}, \ee where scaling
dimensions are given by $[\beta]=-2,~[\varpi]=3,$ and $[c]=2$ for
the $z=3$ HL gravity. {\it This is exactly the same form as the
UV-tensor propagator (\ref{tenprp}) up to $p_0^4$ } when using the
replacement of $\beta\to 1/\omega$ which was derived for entropy
of  the Kehagias-Sfetsos black hole without the Cotton tensor
($C_{ij}=0$)~\cite{Myungent}. However, considering terms beyond
$p_0^4$ ($p_0^5$ and $p_0^6$), we could not make a definite
connection between two propagators even though  highest space
derivative  of sixth order are found in both propagators.
Explicitly, the $p_{0}^5$-term is absent for the GUP-corrected
propagator and coefficients in the front of $p_0^6$ are different.
Two coefficients are the same for  $\eta^4=128/\kappa^2$.

\section{Discussions}
We have explored  a close connection between generalized
uncertainty
  principle (GUP) and deformed Ho\v{r}ava-Lifshitz (HL) gravity.
It was proposed that the GUP commutation relations describe the
UV-quantum theory, while the canonical commutation relations
represent the IR-quantum theory. Inspired by this UV/IR quantum
mechanics,  we obtain the GUP-corrected graviton propagator by
introducing UV-momentum of $p_i=p_{0i}(1+\beta p_{0}^2)$ with
$p_{0i}$ the IR momentum. We compare this with tensor propagators
in the HL gravity. Two are the same up to $p_0^4$-order,  but the
$p_0^5$-term arisen from the crossed term of Cotton and Ricci
tensors did not appear in the GUP-corrected propagators.

Importantly, we confirm that the deformed HL gravity with $\omega$
parameter contains effects of quantum gravity implied by the GUP
with the linear-order of $\beta$ when using a relation of
$\beta=1/\omega$. This means that the deformed $z=2$ HL gravity
without Cotten tensor  could be well described by the
GUP~\cite{ho2}. This  Lagrangian is given by  \be \tilde{{\cal
L}}_{z=2}=\sqrt{g}N
\Bigg[\frac{2}{\kappa^2}\Big(K_{ij}K_{ij}-\lambda
K^2\Big)+\mu^4\Big(R
+\frac{1}{2\omega}\frac{4\lambda-1}{3\lambda-1}R^2-\frac{2}{\omega}R_{ij}R_{ij}\Big)\Bigg].
\ee The tensor propagator is derived from the above  Lagrangian on
the Minkowski background where Ricci-square term $R^2$ does not
contribute to the bilinear term of $t_{ij}t_{ij}$. Hence, it is
easily shown that  $\frac{2}{\omega}p_0^4$-term in the tensor
propagator (\ref{tenprp}) comes from $R_{ij}R_{ij}$-term. On the
other hand, the modified Heisenberg commutation relation
(\ref{UVCR}) is satisfied to linear-order $\beta$ when calculating
the GUP-corrected propagator (\ref{gupcoprop}). Therefore, it is
valid that the deformed $z=2$ HL gravity without Cotten tensor is
well
 explained  by the GUP.

However, it needs  a further study in order to make  a clear
connection between $z=3$ HL gravity and the GUP with second-order
of $\beta$ ($\beta^2$) because the former contains the Cotton
tensor $C_{ij}$ and the replacement (\ref{RPCR}) is obscure.

\section*{Acknowledgement}
This work  was supported by Basic Science Research Program through
the National Science Foundation (KRF)  of  Korea funded by the
Ministry of Education, Science and Technology (2009-0086861).

\end{document}